# Malware Analysis on AI Technique

Anjani Gupta

School of Computer and System Science

Jawaharlal Nehru University

NewDelhi, India

anjani.gupta1197@gmail.com

Dr. Karan Singh

School of Computer and System Science

Jawaharlal Nehru University

NewDelhi, India

karancs12@gmail.com

***Abstract*****—**In today's world, we are performing our maximum work through the Internet, i.e., online payment, data transfer, etc., per day. More than thousands of users are connecting. So, it's essential to provide security to the user. It is necessary to detect and prevent malicious object from gaining persistence and causing destruction within the organization. Therefore, Malware analysis is needed in order to secure the system. This necessitates the use of effective and efficient approaches for detecting OS malware. Due to the cheap cost of technology, artificial intelligence has also become less difficult to implement in projects to analyse malware. The categorization and analysis of malware on OS using various AI-based analysis techniques are covered in detail in this paper.

## 1. Introduction

Malware has arisen as one of the most significant risks to computer systems. It is software intended to drastically alter a computer system, server, or client in favor of leaking sensitive or private data, obtaining unauthorized access to information or discrete values that convey data, and degrading a user's computer security and privacy without them being unaware of it. When the user downloads that file, it unknowingly infects software while clicking an infected link, via email attachment, or by an infected USB. Some forms of malware try to duplicate or replicate. Give an example of a recent malware attack. The "All India Institute of Medical Science (AIIMS)", a premier medical research facility in India, suffered a ransomware attack on November 20, 2022. [1]. It took almost two weeks to get the infected system online again. That record involves approximately 40 million, including some of the most influential person papers, and attackers demand RS 200 crore as a ransom [1].

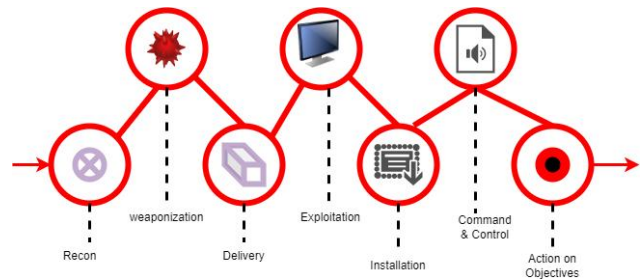

**Figure 1: Attack flow chain for Cyber Security**

To understand the attack process, the concept of attack flow chain is useful in the field of cyber security. It assists organisations in developing proactive security measures, improving incident response capabilities and enhancing overall cyber security posture. Basic steps for attack flow chain are given below in figure 1.

**Reconnaissance**: It is the initial stage of the chain, which is meant to gather as much data as possible on the victim. Possible targets may be located, their weaknesses may be discovered, which third parties are connected to them, and both new and current entrances may be investigated during the reconnaissance phase. It is possible to conduct surveillance both online and offline.

**Weaponization:** After surveillance and the attacker has gathered all pertinent information about potential targets, including vulnerabilities, the weaponization phase of the Cyber Kill Chain begins [19].

**Delivery:** To access users during the delivery phase of a target's network, cyberweapons and other Cyber Kill Chain technologies are employed. Delivery could involve sending users phishing emails with malicious attachments and alluring subject lines [18].

**Exploitation**: During the exploitation step of the Cyber Kill Chain, attackers employ the vulnerabilities they discovered in earlier stages to further enter a target's network and accomplish their objectives[8].

**Installation:** After successfully exploiting a target's vulnerabilities to obtain access to a network, cyber attackers try to install malware and other cyberattacks onto the target network in order to take over its systems and exfiltrate vital data during the installation stage of the Cyber Kill Chain.

**Command and Control**: Cybercriminals develop hacking tools and equipment to interact with malware they have put on a target's network in order to achieve their goals [15].

**Action on Objective**: Cybercriminals create hacking tools, install them on a target's network, and then seize control of that target's network to start the final phase of there objectives [15].

There are numerous kinds of malware, including Adware, rootkits, virus, worm, trojans, ransomware, Keyloggers, etc. [2][7]. Some of them are given below.

**Worm:** It is a type of malware that replicates itself, may duplicate itself without requiring human input.

**Adware:** It display unwanted advertisements on screen, usually while using a web browser [7].

**Virus:** It is a form of malware that can reproduce or duplicate itself and attaches itself to different programs and files. The virus spreads throughout the system when programs or infected files are run on a device [7].

**Rootkit:** Rootkit malware is used by cybercriminals to seize control of a target computer or network. Although it typically consists of a number of tools, it occasionally takes the form of a single piece of software and gives hackers administrative access to the target device [2].

**Ransomware:** Users are unable to access their data or devices, and access is only available in exchange for payment. Ransomware comes in two flavors: Crypto and locker. [3][4]. The system's files are encrypted by crypto-ransomware, rendering them unavailable. File-Locker ransomware is another name for crypto-ransomware. The Locker ransomware prevents users from using their system by locking their desktops or displaying a window that never closes; it does not change data. It also goes by the name "Screen-Locker ransomware" [4].

**Trojan:** It is malware that gets installed into a computer by posing as a useful program. In many cases, the delivery method comprises an attacker hiding dangerous software in legitimate software and using social engineering to trick people into giving them access to the system [2].

**Keylogger:** Some malware can continuously monitor every keyboard and send this information to far-off places where login credentials, including passwords, can be collected and utilized.

The piece is set up as follows. Subsection 2 describes malware analysis and It's types. Subsection 3 covers a literature review, Subsection 4 describes the analysis of malware using AI techniques, and Section 5 offers a conclusion.

## 2. Malware Analysis and It's Types

It is the process of looking at harmful application, also referred to as malware, to understand its behavior, functionality, and potential effects. To determine the function, capabilities, and potential threats of the malware, its code, structure, and characteristics must be studied.

The entire process of malware investigation and categorization is described in Figure 2. We are gathering samples that could be either benign or malicious. After gathering information, a decision is made based on the tools at hand, a data set is prepared, and a model is used to check the correctness and outcome. Alternatively, we may take the report and manually create the dataset or take the data set directly and apply the model.

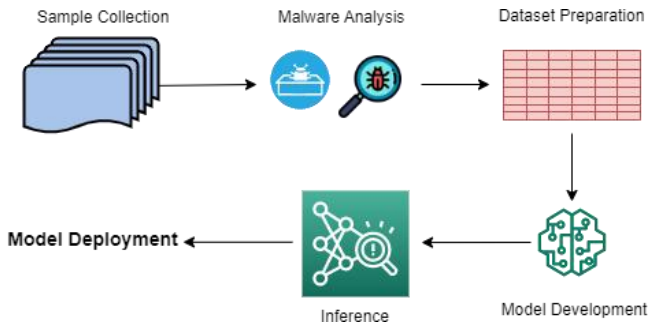

**Figure 2: Process for Analyzing Malware**

Analysis is categorized into three ways: static, dynamic, and hybrid analysis [7].

**Analysis of Static Malware** : It looks at the structure of the executable file without having it execute under supervision. The completed file has a lot of static qualities, such as its many sections and good memory usage. [6].

**Dynamic malware analysis**: Malware is examined in a virtual setting. When malware is activated, it takes control of the privileged mode of the operating system and updates a registry key maliciously to create a virtual environment and protect the host machine. Dynamic analysis software can adjust the computer's registry keys and operate in debugger mode in a virtual environment [6].

**Hybrid Analysis**: It combines static and Dynamic analysis; combining both techniques will provide the best approach. From statically generated and previously unobserved code, hybrid analysis can uncover harmful hidden code and extract many more Indicators of Compromise (IOCs). It is capable of detecting unknown threats, including those brought on by the most sophisticated malware. hence able to identify harmful code.

*2.1 Malware Detection*

It is the process of scanning a computer and its contents for malware. It is effective in identifying malware because it use a variety of techniques and procedures. It's a complicated process that involves both directions. There are different malware detection methods like Signature based, Heuristic**,** Sandbox, Removal tools, Anti- Malware software [2][7].

**Signature-Based:** By employing virus codes, detection can identify malware. It can be recognized mainly to a special code that it carries. When a file enters the computer, the malware scanner collects the code and transfers it to a cloud-based database. There are a lot of viral codes in the database. If the file's code is found in the list, the database will report that the file is malicious. The file is rejected by the anti-malware program, which then deletes it [2].

**Sandbox:** The anti-malware program creates a safe place called a "sandbox" on the computer to store any files that are unfamiliar or suspicious.

**Heuristic**: - Heuristics use formulas to identify malware. It has defined guidelines that files must follow. Several potential rules include: - Illegal to manipulate cameras, no direct access to the hard drive permitted [7].

**Removal Tools:** - It quickly gets rid of the malware. The files and crucial data are entirely secure now that the dangerous file has been removed from the computer.

**Anti-Malware software:** To combat malware, a program named anti-Malware was developed. It protects the machine and makes sure there is no virus by routinely scanning.

*2.2 Malware Classification*

It is the process of categorizing a malware sample into a certain malware family. To create signatures for detection and classification, malware families that share similar traits can be combined. Signatures are either static or dynamic depending on how they are extracted. A binary assembly instruction, a string of bytes, or an imported Dynamic Link Library (DLL) can all serve as the foundation for a static signature. Network connections, file system operations, terminal commands, and function and system call sequences can all be the basis for dynamic signatures.

### 3. Literature Review

*3.1* **Malware Analysis Methods**

- Anderson et al. [16] Machine learning models can be trained using the labeled benchmark dataset EMBER to identify malicious Windows portable executable files statically. The dataset contains characteristics retrieved from 1.1M binary files, including 200K test samples and 900K training examples (300K malicious, 300K benign, and 300K unlabelled) (100K malicious, 100K benign). In addition to the dataset, also make available open-source code that may be used to extract characteristics from additional binaries and add them to the dataset.

- Bosansky et al. [17] To train malware detection models, they can use various datasets. Elastic Malware Benchmark for Empowering Researchers (EMBER) and the more current Sophos/ReversingLabs 20 million sample dataset are two of the most widely used recent datasets (SoReL). The benefit of these datasets is that security professionals have already examined them, and they offer essential feature data that has been extracted and pre- processed for each sample in structured text format. Additionally, several of the examples receive disarmed binaries via SoReL. There are numerous additional internet data sources, some of which even offer samples of actual malicious binaries. If more data on the model has to be extracted, these binaries may be used. SoReL samples cover the same static features as EMBER and have a structure quite similar to that of EMBER. Again, JSON is used to express these. In addition to the over 15M Stores data samples labeled and the additional 4M unlabeled samples, SoReL contains over 10M actual binary malware samples that have been deactivated by modifying their header file [17].

- Ronen et al. [19] Over 600M machines use Microsoft's real-time anti-malware detection tools. It creates millions of data points per day that must be examined for signs of malware. One of the leading causes of these large numbers of separate files is that malware writers make dangerous components polymorphic to avoid detection. It means that hazardous files that are part of the same malware "family" and that exhibit the same types of destructive behavior are frequently altered or disguised to make them seem to be multiple different files. Grouping the files and determining their families is necessary for efficiently analyzing and classifying many such files. Microsoft made a malware dataset of unheard-of magnitude available to the data science and security sectors. It is made up of a collection of recognized malware files from a variety of nine different families. Each malware file has a class label, an integer representing one of the nine family names, and an identifier, a 20- character hash value that uniquely identifies the file.

- Angelo Oliveira et al. [20] Foreign languages, medical diagnosis, overhead photography, and retail products have all been included in their extensions to different fields. The accuracy for various classification tasks usually exceeds 90% for all classes when using contemporary DL and CNN. The current endeavor attempts to evaluate a related structured challenge and assess algorithmic performance compared to recent virus and antivirus detecting methods. Malware-image problem as a well-known MNIST variant, creating nine virus clusters based on byte similarities and identifying the virus family using a greyscale thumbnail image (32 x 32) [20]. A convolutional neural network (CNN) was trained for benchmarking using transfer learning (MobileNetV2) over the course of 50 iterations, with a learning rate of 0.001 and introduced smaller (10,000) and bigger (50,000) instances to assess the effects of malware counts in the dataset. A fast-learning architecture that is sized, accurate, and quick on mobile devices is provided by MobileNetV2 [20].

**Table 1: Malware analysis on different datasets using various methods**

| Dataset name | Method | Year | Features set/ Malware families |
|---|---|---|---|
| EMBER | ROC | 2018 | Parsed features, Format-agnostic features. |
| EMBER, SoReL | PE Ontology | 2023 | Debug, Relocation, Resources, Signature, TLS, CLR, Non-executable Entry Point Exports, Multiple Executable Sections, Symbol, URL Strings, Registry Strings, Path Strings, Non-standard MZ, Low Imports Count [17] |
| Malware | IDA disassembler | 2018 | Simda Ramnit Obfuscator, Lollipop, Kelihos_ver3, Tracur, and Kelihos_ver1. Gatak, ACY [19]. |
| MNIST | K-mean algorithm | 2019 | Portable Executable [20]. |

*3.2 Malware Detection*

- Ijaz et al. [6] proposed techniques that have Registry, API, DLLs, and summary information for dynamic analysis. More than 2300 features are recovered from the dynamic analysis of malware using the Cuckoo sandbox, and 92 elements are removed statically from binary malware using PEFILES. Their suggested method obtains a 94.4% accuracy rate on their dataset, which consists of 800 benign and 2200 malicious files.

- Hu et al. [22] The original space is high-dimensional, and the incoming data are linearly separable. The suggested approach completes the Kaggle Platform's Microsoft Malware Classification Challenge without the need to convert the input vectors to a higher-dimensional feature space. Using information from IDA Pro outputs, executable PE files and PE headers are recreated using hex dumps for PE files that have been converted into raw binary. The total number of viruses is employed for feature extraction, and 40 antivirus products create the results. Malware is taught to recognize malware families.

Malware samples can be classified based on machine instructions and AV software to determine the malware family.

- Mosli et al. [8] various classifiers such as SVM, SGD, DT, RF, KNN, and extracted features are the registry, imported libraries, and API Calls. Out of all implemented ML algorithms, SGD achieved the highest performance in terms of accuracy, 96%. They experimented on 400 malware and 100 benign samples.
- Tian et al. [5] Repacked approach is used to detect malware in android, coupled with partition-based detection, which lowers the rate of false positives, and fuzzy hacking is used to detect fingerprints. Does this outperform the prior approach in terms of results.??
- Carlin et al. [22] It's been proposed that count-based algorithms produce superior results when malware is categorized with labels and detected using count-based and sequence-based algorithms.

### 3.2 Malware Classification

- In Liang et al. [9], The file system, registries, and network artifacts are used to implement SVM, DT, and KNN classifiers.
- They employed a total of 1980 pieces of malware, and accuracy was 95% with a 5% false positive rate.
- Rani et al. [13] perform ransomware family classification using several machine learning methods such as Decision Tree, RF, KNN, SVM, XGBoost, and Linear Regression. They used various features related to API calls, registry keys, files and directory operations, and embedded strings. They utilized the mutual information (MI) score method to select the features from the 30970 feature set and achieved the highest performance of 98.21% with the XGBoost and Linear Regression model.
- Mangialardo et al. [14] Malware analysis methods, both static and dynamic, have shortcomings.However, the experiment yielded better results when the author integrated the two approaches, classifying the malware using the FAMA framework and the random forest algorithm.
- Rani et al. [15] solved the research gap of the limited dataset and model generalization problem for malware classification. They employ a few-shot learning method to solve fixed dataset problems for malware classificationBy turning malware samples into grayscale images and using a variety of Model Agonistic Meta-Learning (MAML) techniques, they take advantage of the visual characteristics of the virus to improve the model's performance against unknown malware. They reported the performance of five diverse datasets and 98.71% accuracy for limited dataset problems and 72.06% for generalized malware classification.

# 4. AI Techniques and it's application on Malware analysis

Artificial Intelligence techniques are increasingly being utilized in malware analysis to enhance the detection and analysis of malicious software. Here are some common AI techniques used in malware analysis:

## 4.1 ML Algorithms

This types of algorithms are divided into: - Supervised, Unsupervised, and Reinforcement learning.

**Supervised learning**: It is trained on a labeled dataset. A labeled dataset means both input and output are available. It is separated into two categories: - Regression and classification.

**Unsupervised learning:** This is known as unsupervised learning when it can supply a set of unlabelled data that must be investigated and has patterns to be found. It is divided into two types: - Clustering and Association.

**Reinforcement learning**: By doing actions and seeing the outcomes, an agent learns how to operate in a given environment via machine learning.

## 4.2 Supervised Learning Algorithms to detect Malware

For analysing and detecting malware, various machine learning is used few are: - Decision Tree, RF, SVM, NB, KNN , SGD and LF [4].

**Decision Tree:** Classification and regression issues can be addressed using the supervised learning technique known as a decision tree. It is a tree-structured classifier, where each leaf node indicates the classification outcome and the interior nodes contain the dataset's features. The two decision tree nodes are Decision Node and Leaf Node. Decision nodes are used to create conclusions and comprise a variety of components, whilst Leaf nodes are the results of decisions and do not include any extra branches [13].

**Random Forest:** A classifier that averages the outcomes of numerous decision trees applied to various dataset subsets in order to raise the expected accuracy of the dataset [11].

**Support Vector Machine:** The SVM approach seeks to identify the ideal line or decision boundary that can divide n-dimensional space into classes in order to quickly classify new data points in the future. SVM is used to choose the extreme vectors and points that contribute to the hyperplane. Support vectors, which are used to represent these extreme circumstances, are the foundation of the SVM technique [4].

**K-Nearest Neighbours (KNN):** This algorithm for supervised machine learning can be used to solve classification or regression issues. Additionally, imputation of missing variables frequently uses it [1]. It is predicated on the idea that an observation's nearest neighbors are those in the data set that are most "like" the observation in question. As a result, we can categorize unknown sites based on the values of the observations that are already made and are the closest to them. By selecting K, the user can specify how many near words will be used in the technique.

Various performance metrics are used to decide the efficiency of a machine learning algorithm [13]. These are Accuracy, Precision and Recall.

*4.3 Malware Detection Method Based on SVM*

It is a method that operates as follows is available for Windows computers. The existing malware's API calls are converted first into a set of vector data. A SVM then learns these vector data as a ransomware feature. The suggested system also recognises unknown malware as a dangerous programme.

The API requests are represented by a vector in the proposed detection technique, which differs from the solutions already in use. To increases the accuracy of malware detection by closely examining API calls the additional machine learning techniques. In other words, SVMs are suitable for identifying malware that is unknown. A SVM learns the vector representation of the log in the proposed malware detection technique is made up of a series of API calls that malware generates as a characteristic of dangerous programmes. The SVM then determines whether an unknown programme is dangerous or benign software for a particular log of the API calls it outputs.

The probability of harmful software being unexplored is reduced, thus an unknown malware is successfully detected. to perform dynamic analysis experiments using actual, widely-used malware in a secure setting that use a sandbox. The study's findings show that the suggested technique increases prediction accuracy and lowers the malware missed rate.

*4.4 Logistic Regression Based Malware Detection*

It provides insights into the relationship between the features and the probability of a sample being malware. However, its effectiveness depends on the quality and relevance of the extracted features and the representativeness of the training dataset.

To create a defence system that employs snort IDS, Logistic Regression, and ANOVA F-Test to prevent polymorphic malware. Snort IDS uses logistic regression with ANOVA F-Test and significant features to find polymorphic malware [12]. Polymorphic malware has been successfully thwarted using this defence arrangement on a windows system that is vulnerable. The ANOVA F-Test significantly increased the detection skills of machine learning models.

## 5. Conclusion

Malware and its relatives, including ransomware, viruses, trojans, and many others, can be analysed in various ways. Because of their effectiveness, accuracy, and resilience, Artificial Intelligent-based solutions are becoming increasingly popular for malware analysis and identification. No technique is 100% error-free. Furthermore, none of the above resolutions can be relied upon to find every family. Most of them succeed in classifying and analyzing some types of malwares but fall short of others. The number of optimal features picked for training determines how effective the machine learning model is. Here, feature engineering is essential for developing a successful analytic system. Uncommonly used elements in malware programmers, such as function calls linked to cryptographic algorithms/file access to significantly modified/deleted [1] data/files, accessing a large number of various directories, etc., can be used to identify malicious or benign[1] code. Techniques like PE (Program Executable), API (Application Programming Interface), and Registry key can be used to determine whether characteristics accurately distinguish between benign and malicious behaviour. However, multi-layered DL models with automatic feature engineering for massive datasets can analyse and classify malware.